\documentstyle[11pt,paspconf,epsf]{article}

\newcommand{\ltsima} {$\; \buildrel < \over \sim \;$}
\newcommand{\gtsima} {$\; \buildrel > \over \sim \;$}
\newcommand{\lta} {\lower.5ex\hbox{\ltsima}}
\newcommand{\gta} {\lower.5ex\hbox{\gtsima}}

\begin{document}

\title{DISK-CORONA MODELS AND X-RAY EMISSION FROM SEYFERT GALAXIES}

\author{Laura Maraschi}
\affil{Osservatorio Astronomico di Brera, via Brera 28, 20121 Milano, Italy}

\author{Francesco Haardt}
\affil{Department of Astronomy \& Astrophysics, Institute of
Theoretical Physics, G\"oteborg University \& Chalmers University
of Technology, 412 96 G\"oteborg, Sweden}

\begin{abstract}
The current status of understanding of the X-ray emission from
Seyfert galaxies involves Comptonization 
of soft photons by hot subrelativistic  electrons. 
After briefly reviewing the early theoretical basis for the 
presence of hot optically thin
plasma in or around accretion disks and the key observations that led to 
develop the presently popular
 model of an accretion disk with a hot corona, we summarize recent progress
in accretion models that take into account energy dissipation and/or angular 
momentum   transport in the corona. Finally, adopting the simple scheme
of a homogeneous plane parallel corona, 
we discuss in detail the dependence of the X-ray spectrum on the coronal 
parameters. Despite the strong coupling between optical depth and temperature
 which determines a spectral shape insensitive to their precise values,
moderate spectral changes are possible. The spectral variability
patterns can be used as diagnostics for coronal physics and should
allow to determine whether the optical depth of the corona is dominated 
by $e^+e^-$ pairs.     
\end{abstract}


\section{Introduction}

The main uncertainty in the thory of accretion disks is on the physical processes
responsible for energy dissipation and angular momentum transport. 
Despite important progresses in understanding turbulent and magnetic viscosity
it is still impossible to anticipate the precise mechanisms at work
in astrophysical plasmas, in particular in accretion disks.
The emitted spectra are not too sensitive to these  mechanisms if the disk
is optically thick and the dissipation occurs within the thick layer.
However the high energy spectra emitted by black hole binaries and active galactic
nuclei indicate conditions far from thermodynamic equilibrium, implying that the
above conditions are  strongly violated in the inner regions of
highly luminous accretion disks around black holes. 

We review here shortly the evolution of accretion models involving a hot corona
as an important site of energy dissipation and the observational grounds for
such picture in the case of Seyfert 1 galaxies.
 We describe the basic features of the radiative coupling and feed
back effects of a sandwich configuration that can account for the 
observed X-ray spectra and present new results on the spectral "dynamics"
expected in such models.  

\section{Early Ideas about X-ray Emission from Accretion Disks}

The importance of Comptonization for the high energy spectra of
 luminous accretion disks
was recognized since the seminal paper  by Shakura and Syunyaev (1973) and
it is a pleasure to recall that  both authors participated to this conference.
However the high temperature and power law spectral shape 
observed for Cyg X-1 and the 
 instability problems of radiation
pressure supported and optically thin disks  challenged the selfconsistency of the
"standard" model  leading  
 Shapiro, Lightman \& Eardley (1976) 
to hypothesize that pressure support could be due to  
hot ions, weakly coupled to the radiating electrons in a "two temperature"
optically thin plasma.
Applying the model to Cyg X-1 these authors noted the desirability of
having the comptonization parameter $y=\tau (\Delta\epsilon/\epsilon)$
close to one, without however finding a motivation for such value within the
model itself ($\tau$ is the optical depth of the corona and
 ($\Delta\epsilon/\epsilon$) the average energy gain per scattering).

The possibility of a hot corona above a relatively cool accretion disk 
was discussed by Liang and Price (1977) 
 and subsequently  Liang (1979) 
first noted  that a sandwich  configuration would
\begin{enumerate}
\item{} be "natural" in a 
magnetoturbulent disk since  dissipation of magnetic energy occurs
preferentially  in a rarefied atmosphere,
\item{} solve the stability
problem (see  Ionson \& Kuperus 1984 ),
\item{} imply a
feedback between disk and corona that would determine a value of $y$
close to 1.
\end{enumerate}

The physics of magnetic dissipation in accretion disks was substantiated
and applied to galactic black hole candidates 
by Galeev, Rosner \& Vaiana (1979), 
 while a basic treatment of
Comptonization was given by  Sunyaev \& Titarchuck (1980).

The relevance of Comptonization for AGN spectra was first discussed by Katz (1976)
and later by Lightman Giacconi and Tananbaum (1978) 
for the X-ray spectrum of NGC 4151.
The latter authors first noted that the inferred parameters for NGC 4151
implied high optical depth for pair production suggesting  a cut off in the
observed spectrum at 500 keV due to this process.

Non thermal pair plasmas were studied in depth in the 80's in relation to
the X-ray emission of AGN . We will not
try to summarize this subject here but refer to  Svensson (1994) for a
comprehensive  review.

\section{Key Observations}

Different "types" of observations were instrumental in leading to the present
widely adopted scheme of a cool optically thick disk embedded in a hot corona.
New observations in the medium-hard X-ray range with the GINGA satellite,
 revealed the signature
of "cold matter" in the X-ray spectra of Seyfert galaxies, detecting
the $K_{\alpha}$ emission line of Fe and a broad hump around 10 keV
as expected for "reflected" X-rays (Pounds et al. 1990).

 Systematic comparisons of the UV and X-ray emission in quasars and Seyfert galaxies
(Sanders et al. 1989, Walter and Fink 1993) indicated that the power emitted 
in the
latter was at most comparable to but not larger than that emitted in the UV. 
In order to have similar accretion power, the two emitting regions must be at a
similar distance from the black hole 
and a configuration with a hot inner region surrounded by a larger cooler
disk is problematic.

 A possible way out of this difficulty is provided 
by the new class of advection dominated accretion disk solutions
(see the review by Narayan at this meeting) in which the radiative efficiency of
the inner disk is reduced. In these solutions however, as in the earlier model by
Shapiro Lightman \& Eardley the value of the Comptonization parameter is determined
"a posteriori".
 
 Moreover  observations of correlated 
variability in the UV and X-rays (NGC 4151, Perola et al., 1986, and NGC 5548,
 Clavel et al., 1992) suggested that a large fraction of the UV emission
is due to  reprocessing  of X-rays, implying that the UV emitting region
intercepts a large fraction of the X-rays .

The most natural configuration accounting for the above points is that of a two phase
or sandwich disk model as described below. An important prediction of this
scheme was that a thermal cut-off should be expected above 100 keV in the X-ray
spectra of Seyfert nuclei. 
The OSSE experiment on the Compton Gamma Ray Observatory indeed confirmed this
prediction, showing a substantial steepening of the spectra in the
   hard X-ray range (e.g. Maisack et al. 1993,
Zdziarski et al. 1995). These results set strong limits to the possible 
 role of non-thermal pair plasmas and provide a measurement of the total luminosity
in the comptonized component (due to the "convergence" of the integral flux)
which confirms that the  luminosity in the UV is comparable or larger than in 
X-rays.

\section{A Homogeneous Two Phase Model for the     
 X-ray Emission}


Guided by the  above theoretical and observational points 
we analized the radiative transfer and coupling
in the simple scheme of a sandwich model with a cool optically thick layer 
 embedded in a hot optically thin one.
Using a plane parallel limit a thermal balance equation can be written for 
each of the two "phases" assuming that a fraction $f$ of the gravitational
power is dissipated in the hot corona and a fraction $1-f$ in the optically
thick disk.
The coupling between the two "phases" arises from the fact that the Comptonization 
process in the hot corona gives rise to quasi-isotropic X-ray radiation.
Therefore, in the plane parallel limit, only half of it escapes from the source 
while the other half impinges on the cool disk. The latter is in part (10 -- 20\%)
 reflected,
giving rise to the observed spectral hump in the 10 -- 30 keV range; in small part 
it is reemitted as an Fe fluorescence line, but the largest part (80 -- 90 \%)
 is absorbed,
 reprocessed and reemitted into black-body photons which contribute  to
 the soft photon input for Comptonization.

The two coupled equations determine the value of the Compton amplification parameter
$A = 1 + L_{Comp}/L_{soft}$, necessary to close the loop (Haardt \&
Maraschi 1991, hereinafter HM91). $A$ is determined by the only free parameter
$f$ and for small f $A\simeq 1+f$, while for $f=1$ $A\simeq 3$. The exact
value of $A$ in the limit $f=1$ depends on the precise value of the 
albedo and on the degree of anisotropy of the comptonized radiation and is very
important because it fixes the spectral shape of the comptonized X-ray radiation.
In the theory of Comptonization the spectral index depends on the temperature $T_C$
and optical depth of the comptonizing electrons. Therefore fixing $A$ 
implies a fixed spectral shape and a relation between $T_C$ and $\tau$.

If the corona is dominated by electron positron pairs
its optical depth is determined by the compactness parameter, defined as
 $l_c ={ {\sigma_T}\over{m_e c^2}} H/R^2 L$ where H is the scale height of the corona,
R the radius of the underlying disk and L the total luminosity of the system.
The relation between $T_C$ and $\tau$ imposed by the feed back effect
transforms then into a relation between $T_C$ and $l_c$, with higher 
values of the compactness corresponding to lower values of the 
temperature (HM91).  

From the approximate estimates in HM91 it was clear that in order to reproduce
 the average
X-ray spectrum of Seyfert galaxies {\it it was necessary to have $f\simeq1$}, which
implied that practically all the energy was dissipated in the corona and the
cool disk acted only as a "passive" reprocessor. With this assumption, the model could 
account for the average value and for  the relatively small
dispersion  of spectral indices of Seyfert I galaxies
(Nandra \& Pounds 1994), for the reflection hump and for the Fe emission line.
It  predicted a  spectral cut off in the hard X-ray range which was
 revealed by OSSE (see above).

Improved treatment of radiative transfer and detailed spectral calculations
 (Haardt, 1993; Haardt and Maraschi 1993; Titarchuk 1994) 
confirmed and refined these results. 
More accurate solutions for the pair corona model are discussed in Poutanen and 
Svensson (1996) and Stern et al. (1995), not only for a slab but also for 
cylindrical and hemispherical geometry for the active region.

\section{Accretion models of disk+corona systems}

The success of the two phase model in explaining the X-ray properties of Seyfert
galaxies and the realization that most of the  accretion power
should be released directly in the corona, presumably through magnetic buoyancy and
reconnection above the optically thick disk, motivated new investigations
of this configuration  with different assumptions  as to the
share of angular momentum tranfer, energy dissipation
 and accretion rate  and with different boundary conditions  between disk and corona.

 Nakamura \& Osaki (1993)  
 describe the two phases as two $\alpha$ disks with accretion taking place 
in the cool disk only. However, as additional constraint
 the average coronal pressure $P_c$ is taken to be equal to the average disk pressure
 $P_d$, clearly an unphysical assumption.
 A similar approach  is adopted by Kusunose \& Mineshige (1994)
who assume that mass accretion and energy dissipation are shared between disk and
corona again with the unphysical condition $P_c=P_d$.

Svensson \& Zdziarski (1994) model the cool phase as an $\alpha$ disk except
for assuming that a fraction f of the power release associated with the angular
momentum transfer takes place above an optical depth $\tau_c$ which is treated
as a free parameter due to the poor knowledge of the microphysical processes.
They show that in the case of a strong corona (large $f$) the  unstable
radiation dominated region disappears (see also Ionson and Kuperus, 1984 and
Chen 1995) 
coronal pressure and irradiation only affects  the surface layers of the cold
disk. 

A more ambitious approach is that of  Zycki et al. (1995) and Witt, Czerny and
Zycky (1996) who  aim at determining the fraction of power dissipated in the
corona selfconsistently rather than treating it as a free parameter. They
describe the disk and corona as accretion disks with different $\alpha$
parameters, imposing pressure balance at the base of the corona, and allowing
the fraction of mass accretion through the disk and corona to be a function of
radius. 

 In the first paper they find that coronal solutions are
not allowed at high accretion rates due to the impossibility of satisfying the
hydrostatic equilibrium equation, while in the second they include vertical
supersonic outflow. In the latter case the corona forms only for accretion rates
larger than a limiting value and the fraction of energy dissipated in the
corona decreases with increasing accretion rate. The interest of this result
is clear in view of the possible comparison with observational results
for  black hole binaries in different intensity states 
or for a comparison of the average properties of Seyfert galaxies and quasars.
Although a large uncertainty still exists in the applicability of an $\alpha$
description of viscosity, especially in this complex case, this approach is
very promising.

\section{The Two--Phase Model Revisited: A Patchy Corona?}

The simple two phase model originally proposed has two strong implications.
First, the luminosity in the soft blackbody photons emitted by the 
surface underlying the 
comptonizing corona must be equal  to the X-ray luminosity 
(neglecting factors of a 
few due to the different angular dpendences of the two components).
Moreover their intensities must be "perfectly" correlated   
in time. If we identify the blackbody component with the total
UV emission, both these predictions are violated.
In fact, as mentioned earlier (see section 3), if
the spectral steepening in the hard X-ray range measured by OSSE
for a few Seyfert galaxies is  a general property,
the estimated "bolometric correction"  for the X-ray luminosity
yields values smaller than the UV luminosity (Walter and Fink 1993).
At the same time the correlation between UV and X-ray luminosity
is present on  timescales of days but  may not hold on shorter
timescales.

Both these difficulties can be solved in a more realistic coronal model,
as proposed by Galeev Rosner \& Vaiana (1979),
 where the energy stored in the magnetic field
is released in  active regions {\it covering only part of the
cool disk} (Haardt, Maraschi \& Ghisellini 1994). 
The feed back and X-ray spectral properties discussed earlier can be maintained
if the power reprocessed under the active regions dominates that
which is locally dissipated by other processes. However the 
 disk emission not covered by active regions can leak out   freely.
The condition that f=1 is therefore not necessary on the large scale.

The active regions should produce hot spots on the disk giving
rise to a rapidly varying component contributing in part to the UV bump
and in part to the so called "soft excess" observed in many Seyferts 
(e.g. Mushotzky, Done \& Pounds, 1993).
Also the flat limit may not be valid for each region
allowing some more freedom in the expected spectral shapes (see for a review
Svensson 1996)

\begin{figure}
\plotone{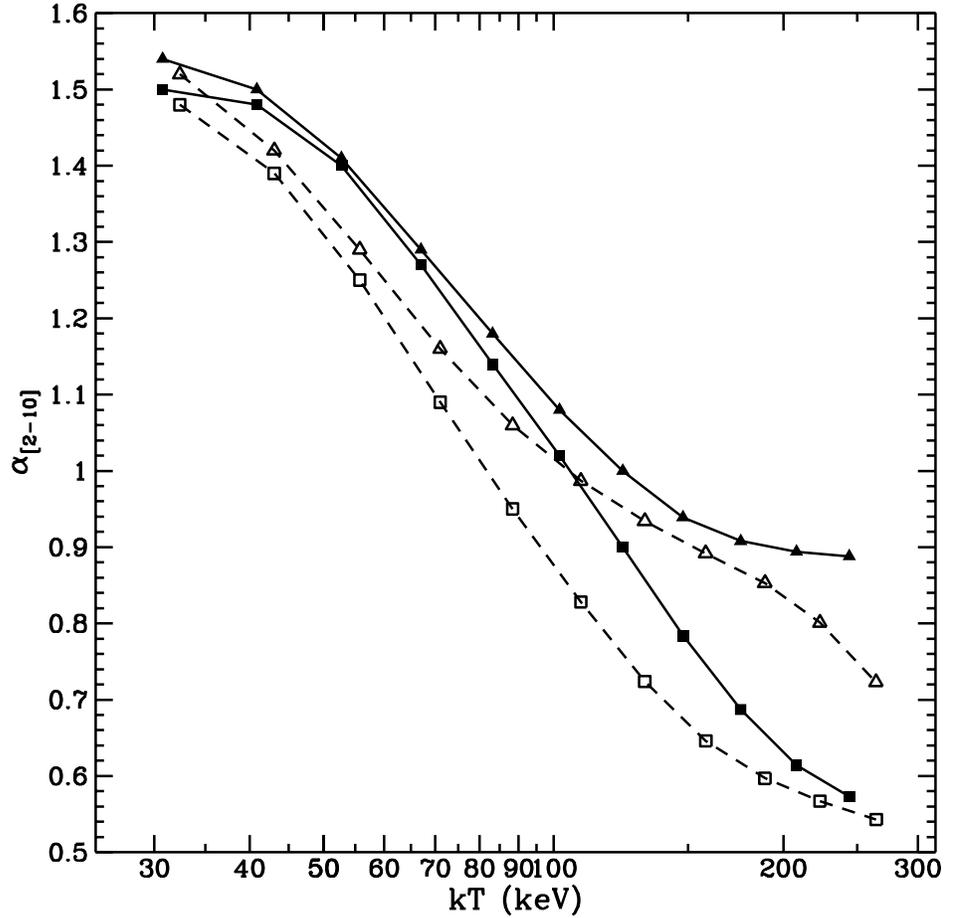}
\caption{The total (direct power law + Comptonized reflection)
spectral index in the [2--10] keV range $\alpha_{[2-10]}$ is plotted against
the coronal electron temperature $kT$, for different values of the viewing
angle and for different temperatures of the soft blackbody photons.
 Square symbols refer to a face--on line of sight, triangles to an inclination
of $60^o$. 
 Open symbols, connected with dashed 
lines, refer to a soft photon temperature $kT_{BB}=100$ eV. Closed symbols 
connecetd with solid lines are for $kT_{BB}=50$ eV. The curves show that
 harder spectra are expected for larger temperatures (lower optical depth) 
of the comptonizing electrons. The difference between 
results computed for different inclination angles is larger at higher 
temperatures (lower optical depths) due to the larger anisotropy of the first
scattering.} 
\end{figure}

\section {Diagnostics of Coronal Models: Broad Band X-ray Variability}

 In view of some existing and many more expected
data on the spectral variability of Seyfert galaxies from the soft to the hard X-ray
band we explored in some detail the behaviours expected in the model
(Haardt Maraschi \& Ghisellini 1996). In fact,
even adopting a fixed geometry for the active region(s) 
and despite an almost constant Compton parameter imposed by the feed back effect,
($L_C/L_S=2$ for the slab geometry adopted in the following) 
the spectral shape in the medium X-ray range (characterized by a spectral index
$\alpha_{[2-10]}$)  depends somewhat on the parameters of the
system (i.e. $\tau$, $T_C$ and $T_{BB}$). 
Moreover, the optical depth may depend on luminosity as well as the temperature
of the soft reprocessed photons. Since $T_{BB}$,  $\alpha_{[2-10]}$ and
$T_C$ can be simultaneosly observed by broad band satellites like XTE or SAX
or by the joint use of different satellites, it is interesting to
compute the relations between these quantities.
The most important results are  summarized below.

The relation between the spectral index of the {\it
total} spectrum  $\alpha_{[2-10]}$ 
(including also Comptonization of the reflected spectrum )
and the  temperature of the corona is shown in Figure 1, for different
values of $T_{BB}$ and of the inclination angle.
We recall that the optical depth  does not appear explicitly but is related to the
temperature by the balance equation for the corona (see also Pietrini and Krolik 1995).
 Values of $T_C$ in the range 30--300 keV, as suggested by observations
 (Maisak et al. 1993, Johnson et al. 1993, Madejski et al. 1995), correspond to
optical depths between 1 and 0.1. 

\begin{figure}
\plotone{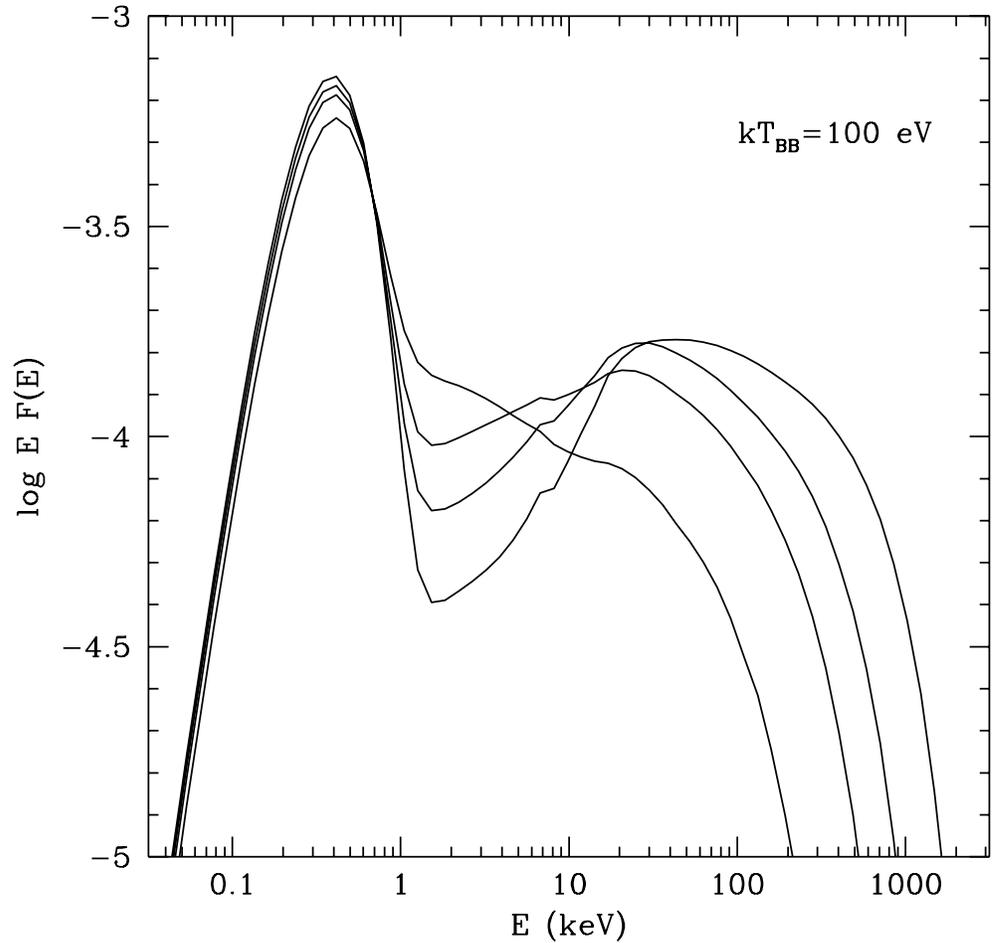}
\caption{Comptonized spectra for constant luminosity and black body 
temperature ($kT_{BB}=100$ eV), computed for different $(\tau,\Theta)$ 
equlibrium values [(0.63,011), (0.32,0.21), (0.2,0.3), (0.1,0.5)]. A 
face--on line of sight is considered.}
\end{figure}

\begin{figure}
\plotone{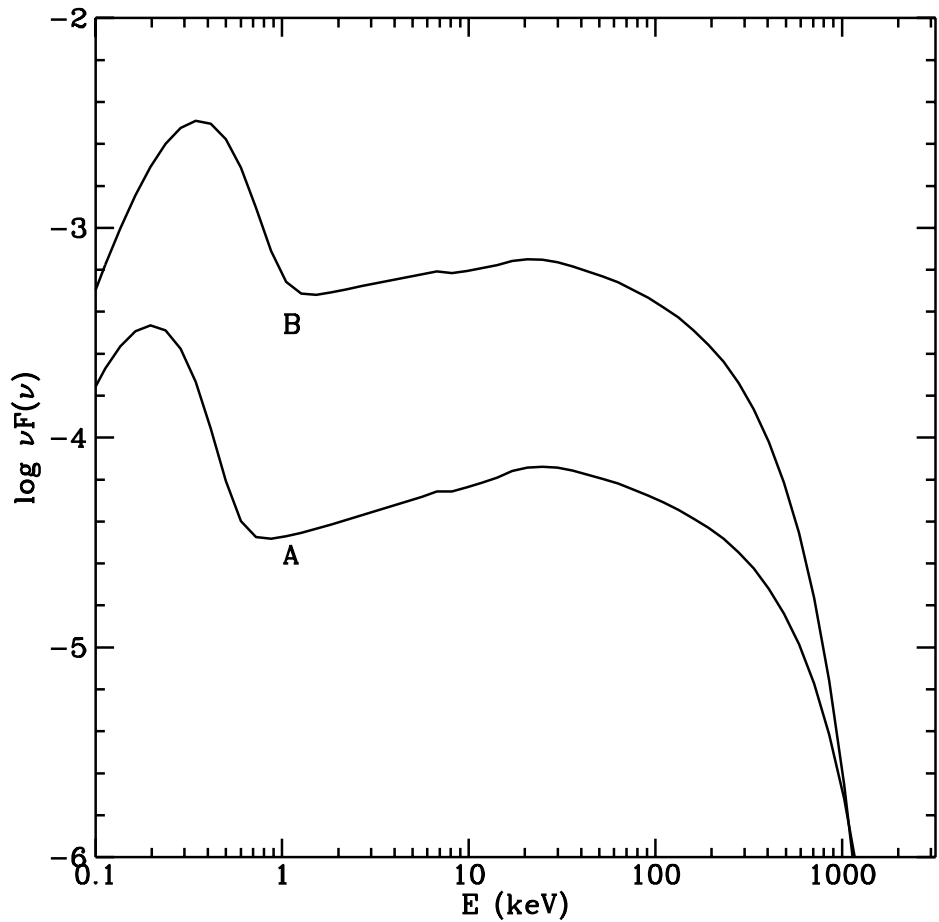}
\caption{Spectra for $\ell_c=30$ (A) and $\ell_c=300$ (B) are computed 
assuming a change of luminosity with constant emitting area. 
Spectrum "A" has $\tau=0.2$, $\Theta=0.3$ and $kT_{BB}=50$ eV. Spectrum 
"B" has $\tau=0.32$, $\Theta=0.21$ and $kT_{BB}=89$ eV. In both cases 
a face--on line of sight is considered.}
\end{figure}

\begin{figure}
\plotone{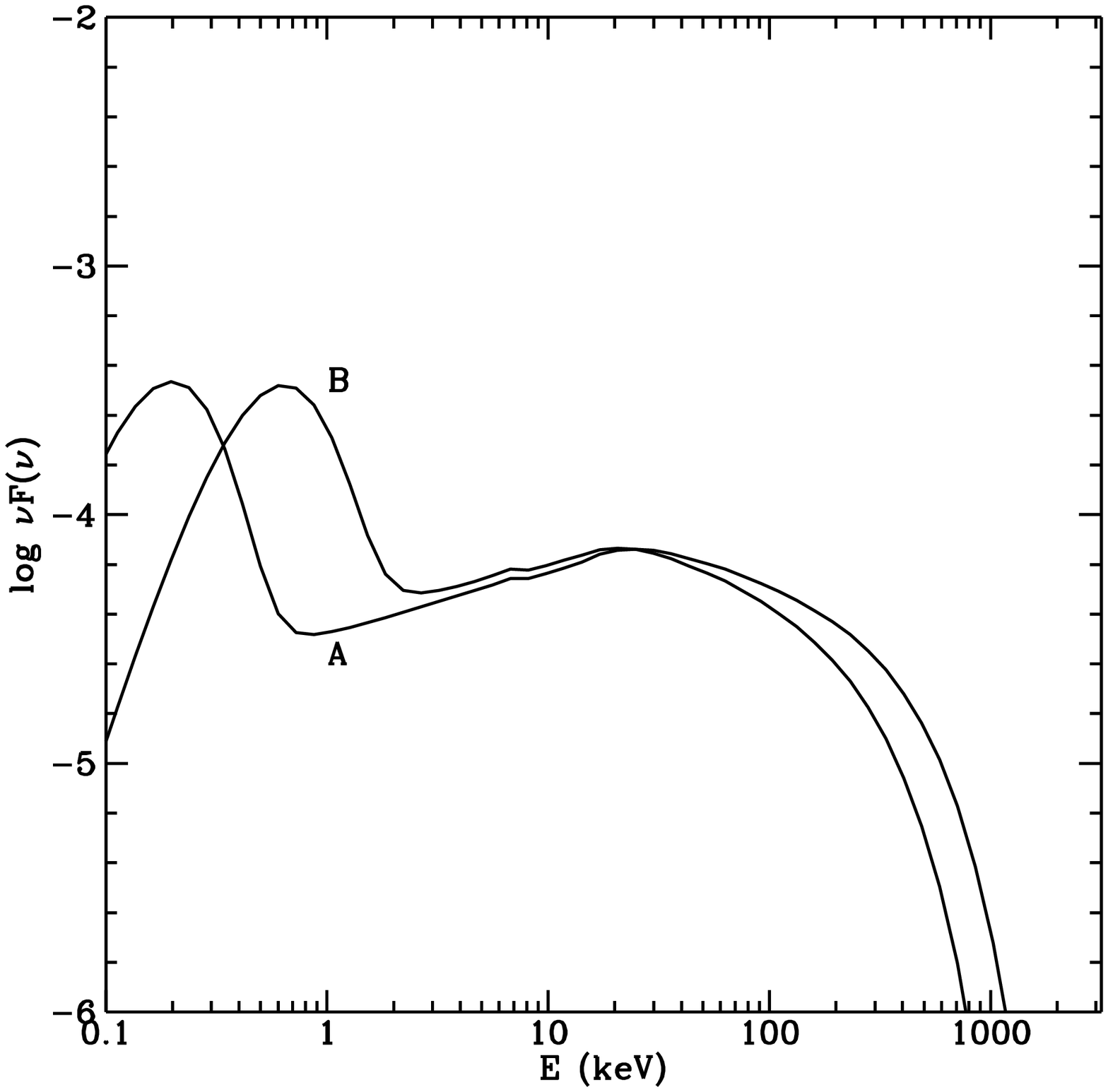}
\caption{Same as Figure 3, but assuming a constant luminosity and a 
varying emitting area. As before, spectrum "A" ($\ell_cv=30$) has 
$\tau=0.2$, $\Theta=0.3$ and $kT_{BB}=50$ eV. Spectrum "B" 
($ell_c=300$) has $\tau=0.32$, $\Theta=0.21$, but a much 
larger $kT_{BB}=158$ eV.}
\end{figure} 

The important points to note are the following: 

\begin{enumerate}
\item{}Changes  in $\tau$ and $T_c$ give rise to significant spectral variability
despite the fact that the ratio $L_C/L_S$ is constant in the model. In the
slab geometry considered here $\alpha_{[2-10]}$ is found to be in the
range [0.4--1.4] for $\tau$ varying between 0.1 and 1. 
\item{} Steeper spectra are produced as $\tau$ increases approaching unity. 
Correspondingly, an increase of the spectral index is accompanied by a  decrease of
the coronal temperature, i.e. {\it spectral index and coronal temperature are
anticorrelated}. The actual values of $\alpha$ and $T_C$ in Fig 2 refer to
$L_C/L_S \simeq 2$, however the anticorrelation between  spectral index and the
electron temperature is a general feature common to models based on Compton
cooling with fixed geometry. 
Note that the dispersion of the $\alpha_{[2-10]}$ vs. $kT$ relation is larger at
higher temperature. 
\end{enumerate}

If the main contribution to $\tau$ comes from ordinary electrons,
a reliable relation between luminosity and optical depth for the corona 
cannot be specified a priori. It is conceivable that changes in $\tau$ 
(e.g. fluctuations) occur at constant luminosity. This case is
illustrated in Figure 2, where broad band spectra of constant total luminosity
and different $\tau$ and $T_C$ are compared. The spectra pivot around
10-30 keV: in the medium X-ray range higher intensity corresponds to 
softer spectra. while in the hard X-ray range larger intensity
corresponds to harder spectra.
 
In contrast to the above situation, if the optical depth is due to pairs
as expected in cases of high compactness, the luminosity (for fixed size)
is a predictable and very strong function of $\tau$. This case is illustrated in
Figure 3, where the two spectra refer to a change in luminosity by a factor 10,
but have very similar spectral shapes. The higher luminosity spectrum also has
a consistently higher $T_{BB}$.

Finally in Figure 4 two spectra due to pair dominated coronae are compared,
where the compactness was varied by changing the surface area of the active
region at constant luminosity. 
Here a dramatic change in $T_{BB}$ is apparent with little spectral variations in the 
medium hard X-ray range. 

The two spectra  also represent an extreme case of a large variation in the soft
band {\it without} a correspondingly large variation at medium X-ray energies.
This demonstrates clearly  that the correlation between $L_c$ and $L_s$
does not translate necessarily into a correlation of the {\it observed}
medium and soft X-ray fluxes. 
 
It seems quite premature to derive firm conclusions from the few broad band
observations presently available. We note however that some spectral 
variability episodes observed in NGC 4151 (Perola et al. 1986),
 NGC 5548 (Done et al. 1995) and Mrk 766 (Leighly et al 1996), resemble more
the pivoting behaviour illustrated in Figure 2 rather than those depicted
in Figure 3 or 4, suggesting that these sources are not pair dominated.

\section {Conclusions}

Substantial progress in understanding emission processes
from the inner regions of accretion disks has been achieved.
 X-ray observations suggest a two phase model, with hot optically thin gas
coupled to a cooler optically thick layer, as may be expected for an accretion
disk with large energy dissipation in a hot corona. However, the formation,
heating and structure of the corona, its dependence on fundamental parameters
like the accretion rate and central black hole mass are still largely unknown.

Broad band measurements from the soft to the hard X-ray band, possible with  
present satellites like ROSAT, ASCA, XTE and SAX offer the exciting perspective
of investigating  how the coronal parameters
vary  in time for the same source or differ statistically in different classes 
of sources.  
Important questions such as:
is the optical depth dominated by $e^+e^-$ pairs? are the coronal parameters related 
to luminosity? what is the cause of  the different X-ray properties of
Seyfert galaxies and Quasars? could be answered in the near future.

\end{document}